\documentclass[11pt]{iopart}
\input{amssym}
\usepackage{epsfig}
\global\arraycolsep=1pt

\newcommand{\be}{\begin{equation}}

\newcommand{\ee}{\end{equation}}
\newcommand{\bea}{\begin{eqnarray}}
\newcommand{\eea}{\end{eqnarray}}
\newcommand{\ben}{\begin{equation*}}
\newcommand{\een}{\end{equation*}}
\newcommand{\bean}{\begin{eqnarray*}}
\newcommand{\eean}{\end{eqnarray*}}

\begin{document}
\title{Macroscopic tunneling, decoherence and noise-induced activation}
\author{Fernando C. Lombardo, Diana Monteoliva and Paula I. Villar}
\address{Departamento de F\'\i sica {\it Juan Jos\'e
Giambiagi}, Facultad de Ciencias Exactas y Naturales, UBA;
Ciudad Universitaria, Pabell\' on I, 1428 Buenos Aires, Argentina}


\begin{abstract}
We study the effects of the environment at zero temperature on tunneling in an open system
described by a static  double-well potential. We show that the evolution of the
system in an initial Schr\"odinger cat state, can be summarized in
terms of three main physical phenomena, namely decoherence, quantum tunneling
and noise-induced activation. Using
large-scale numerical simulations, we obtain a detailed picture of the
main stages of the evolution and of the relevant dynamical processes.
\end{abstract}



\newcommand{\beq}{\begin{equation}}
\newcommand{\eeq}{\end{equation}}
\newcommand{\dalam}{\nabla^2-\partial_t^2}
\newcommand{\mbf}{\mathbf}
\newcommand{\itm}{\mathit}
\newcommand{\beqa}{\begin{eqnarray}}
\newcommand{\eeqa}{\end{eqnarray}}


The tunneling of a particle through a potential barrier is a
fundamental effect in quantum mechanics. Macroscopic quantum
tunneling can be associated with the tunneling of a many-body
wavefunction through a potential barrier, and therefore provides a
more stringent test of the validity of quantum mechanics than one
particle case. One place where the study of macroscopic quantum
tunneling is experimentally accesible is in the tunneling of a
Bose-Einstein condensate (BEC) out of an optical trap. These systems
have a controllable number of atoms and hence straddle
 the boundary between microscopic and macroscopic, and hence between quantum and
classical systems. Proposals to perform interference experiments
using confined atoms \cite{Shin04} rely on the separation and merger
of two potential wells to split and recombine atomic wave packets
\cite{Anderson02}. BECs in a two-well potential have been created in
experiment \cite{Shin05}. Atom-atom interactions tend to localize
particles in either potential well and reduce the coherence of the
splitting and recombination processes \cite{Menotti01}, whereas
tunneling serves to delocalize the atomic wave packets and keeps a
well-defined relative phase between the potential wells.

Macroscopic systems are generally open systems, interacting with an external
environment, and in this context quantum tunneling  \cite{caleg,takagi}
is qualitatively different from its experimentally verified microscopic analogue \cite{hanggi}.
The analysis of open systems has led to interesting results, detailing
the dynamics of a quantum system coupled to a thermal
bath with arbitrary temperature. A closed quantum system described by a
state localized around a meta-stable minimum, should tunnel through the
potential barrier with a well defined time-scale. This tunneling time can be
estimated using standard techniques such as the instaton method
\cite{Coleman}. For an open system, on the other hand, it is well known
that the environment induces decoherence on the quantum particle,
its behavior becoming classical as soon as interference terms are destroyed
by the external noise. This transition from a quantum to a classical
behavior is forced by the interaction with a  robust environment and takes place
at a given time-scale, the decoherence time. This quantity depends on
the properties of the system, its environment and their mutual coupling.
If the decoherence time is
significantly smaller than the tunneling time, one would expect that
after classicalization the state should become confined to the meta-stable
vacuum (or potential well), with tunneling being suppressed. The particle could still cross
the barrier but only if excited by the bath, its energy increasing via
thermal activation, for example. This process is distinct from
quantum tunneling, is classical in its nature and should be efficient
mostly at high environmental temperatures. An interesting question arises: what is
the effect on tunneling if
the particle is coupled to a reservoir at zero temperature?
Though in this case there should be in principle no thermal activation,
we know there is decoherence induced by a quantum environment at zero
temperature \cite{pau}. This would lead to
classicalization and one could conclude that even at $T=0$, quantum tunneling should
be inhibited by the interaction with the external environment \cite{calzetta1}.

The study of the effects of an external environment on tunneling was
initiated in Refs. \cite{caleg}. It was shown that dissipation
inhibits tunneling. Authors consider a two level limit of a particle
in a double-potential well with Hamiltonian $H = -\frac{1}{2}\hbar
\Delta_0 \sigma_x + \frac{1}{2}\epsilon \sigma_z$, where $\epsilon$
is the detuning frequency. In
the case $\epsilon = 0$, the eigenstates can be writen as
combinations of antisymmetric and symmetric states: $\psi_0 =
1/\sqrt{2} (\psi_R - \psi_L)$ and $\psi_E = 1/\sqrt{2} (\psi_R +
\psi_L)$, respectively. Therefore, the probability to stay on the
right or left well is given by an oscillatory function: $P(t) = P_R
- P_L = \cos(\Delta_0 t)$. This result has no classical analog, and
this is the genuine expression of the phase coherence between the
states (i.e. NH3 inversion coherence). In the case this two level
system is coupled to a reservoir (spin-boson model), the tunneling
is inhibited by the dissipation in the limit of $\Delta_0/\Lambda
\ll 1$ ($\Lambda$ is the Debye frequency of the environment) and
$k_B T/\hbar\Lambda \ll 1$ (low temperature limit) \cite{caleg}.

In Ref. \cite{pre} we have studied a general
tunneling  system described by a static Hamiltonian. Specifically,
we have looked in detail at the evolution of a particle in a quantum
state localized at one of the minima of a double well potential,
when coupled to an external environment at both zero and high
temperature. We have presented analytical descriptions of the
effects of dissipation and diffusion, and estimated the time-scales
associated with the distinct physical processes governing the
dynamics of the system: decoherence, quantum tunneling and
activation. A very interesting extension of previous analysis is to
study superpositions of macroscopic quantum states. This is a
crucial aspect when analyzing BECs in double-well traps \cite{frazer}. Therefore,
our goal here is to extend our previous considerations about
decoherence, tunneling and noise-induced activation to the case of
an initial coherent superposition of two Gaussian wave packets. 
The interactions between atoms in finite size affect the
coherence and the relative phase undergoes diffusion due to the
presence of a condensate self-interaction and also the interaction
between condensate and non-condensate atoms creates decoherence \cite{pitaevskii}. In
the experiments on BECs of dilute alkali-metal atomic gases, trapped
atoms are evaporatively cooled and exchange particles with
their environment. Macroscopic quantun coherence of BECs results in
coherent quantum tunneling of atoms between the two modes, analogously 
to the coherent tunneling of Cooper
pairs in a Josephson junction. Thus, our motivation is to see how
robust is the quantum coherence between BECs in the presence of the
environment at zero temperature, and how the tunneling process is
affected. It is of big importance the study of the transition from a
coherent to an incoherent regime associated with the increase of
fluctuations of the relative phase between condensates confined in
each of the different traps.

We will start by considering an anharmonic trap given by
$V(x)=-\frac{1}{4}\Omega^2 x^2 + \lambda x^4$. This is a double well
potential with two absolute minima at $x_0=\pm
\Omega/\sqrt{8\lambda}$
 separated by a potential barrier with height
$V_0=\Omega^4/(64\lambda)$. We will assume that the system (BEC) is
open, meaning that it is coupled to an environment composed of an
infinite set of harmonic oscillators, by which we
model the interaction with the non-condensate atoms.

The dynamics of the non-linear potential can be obtained by tracing
over the degrees of freedom of the environment and obtaining a
master equation for the reduced density matrix of the system,
$\rho_{\rm r}(t)$. We will assume that the initial states of the
system and environment are uncorrelated, with the latter being in
thermal equilibrium at zero temperature for $t=0$. Only when the interaction is
turned on the system is allowed to evolve and the initial condition
is not an equilibrium state. As we are interested in studying
tunneling-like phenomena, we will
 estimate the tunneling time $\tau$ by
looking at the evolution of a state for which the particle is initially
localized in one of the sides of the double potential well. 
Numerically, we found that in general $\tau=3./(E_0-E_E)$, where
$E_E$ and $E_0$ are the energies of the symmetric/anti-symmetric
eigenstates respectively \cite{pre}. The energy difference and
corresponding tunneling time can be obtained by a straightforward
instanton calculation \cite{Coleman}, the final result being
$\tau=\frac{3.}{E_0-E_E}=
\frac{3}{8}\sqrt{\frac{\pi}{2}\frac{\Omega}{V_0}} \frac{1}{\Omega}
\exp{\left[\frac{16}{3}\frac{V_0}{\Omega}\right]}$.
 The expression inside the exponential is
 the classical action for the instanton, $S_0=(16/3)\times V_0/\Omega$.

For the open case, we focus on Ohmic environments with spectral
density $I(\omega)= \frac{2}{\pi} \gamma_0 \omega
\frac{\Lambda^2}{\Lambda^2+\omega^2}$ at $T=0$ ($\gamma_0$ is the dissipation 
constant and $\Lambda$ the frequency cutoff). After a rather
lengthy calculation, the master equation at $T=0$ on the basis of
eigenstates of the isolated system can be written as \cite{pre}
\begin{eqnarray}
\nonumber
&\dot \rho_{\mu\nu}& = - i \, \Delta_{\mu\nu} \rho_{\mu\nu} - \sum_{\alpha\beta}
\{ D_{\alpha\beta} x_{\mu\alpha} x_{\alpha\beta} \rho_{\beta\nu}-
   D_{\beta\nu} x_{\mu\alpha} x_{\beta\nu} \rho_{\alpha\beta}-\\
   \nonumber
  &-& D_{\mu\alpha} x_{\mu\alpha} x_{\beta\nu} \rho_{\alpha\beta}+
   D_{\alpha\beta} x_{\alpha\beta} x_{\beta\nu} \rho_{\mu\alpha}\} + i \sum_{\alpha\beta}
\{ \gamma_{\alpha\beta} x_{\mu\alpha} x_{\alpha\beta} \rho_{\beta\nu} + \\
&+& \gamma_{\beta\nu} x_{\mu\alpha} x_{\beta\nu} \rho_{\alpha\beta}-
  - \gamma_{\mu\alpha} x_{\mu\alpha} x_{\beta\nu} \rho_{\alpha\beta}-
   \gamma_{\alpha\beta} x_{\alpha\beta} x_{\beta\nu} \rho_{\mu\alpha}\},
\label{masterT=0}
\end{eqnarray}
where the time dependent complex coefficients $D_{\alpha\beta} =D_{\alpha\beta}(t)$
and $\gamma_{\alpha\beta} =\gamma_{\alpha\beta}(t)$ are given by $
D_{\alpha\beta} = D(\Delta_{\alpha\beta}) + i\; \Delta_{\alpha\beta} \;f(\Delta_{\alpha\beta})$
and $\gamma_{\alpha\beta} = -\frac{1}{2} \tilde\Omega^2(\Delta_{\alpha\beta}) - i
\; \Delta_{\alpha\beta}\; \gamma(\Delta_{\alpha\beta})$. In this expressions,
$\Delta_{\alpha \beta}=\omega_\alpha -\omega_\beta$, is the frequency
difference between eigenstates $\alpha$ and $\beta$.
The set of coefficients $D_{\alpha\beta}$ encapsulates the effects of diffusion
 at $T=0$, with $D(\Delta)$ representing the normal diffusion and $f(\Delta)$
the anomalous one.
The others coefficients represent the effect of the environment through
the dissipation kernel $\eta$, with $\tilde\Omega(\Delta)$
the frequency shift, and $\gamma(\Delta)$ the dissipation coefficient
(see analytical expressions in \cite{pau,pre}).
We have numerically solved equation~(\ref{masterT=0}), in
the under-damped case, using a
standard adaptative step-size fifth order
Runge-Kutta method for different parameters of the  system and
the environment.
All results were found to be robust under changes on the parameters
of the integration method. From the solution of (\ref{masterT=0}) we will 
show the Wigner function and the probability distributions of the main system 
at any time \cite{diana}.

As an example we have chosen $\Omega=100$ and $V_0 = 200$ for the
system for which the estimated tunneling time scale is $\tau\approx
158.27$. We set the frequency cutoff to
$\Lambda =10 V_0= 2000$. With
 this set of parameters, the effects of the
frequency shift in the initial state can be neglected \cite{pre}. 
The decoherence time $t_D$ has been analytically evaluated in \cite{pau}, getting a bounded
value $t_D \leq 1/8\gamma_0$ for very flat potentials, and $t_D \sim 1/\Lambda$ for 
large values of $\Omega$.

\begin{figure}
\centerline{\psfig{figure=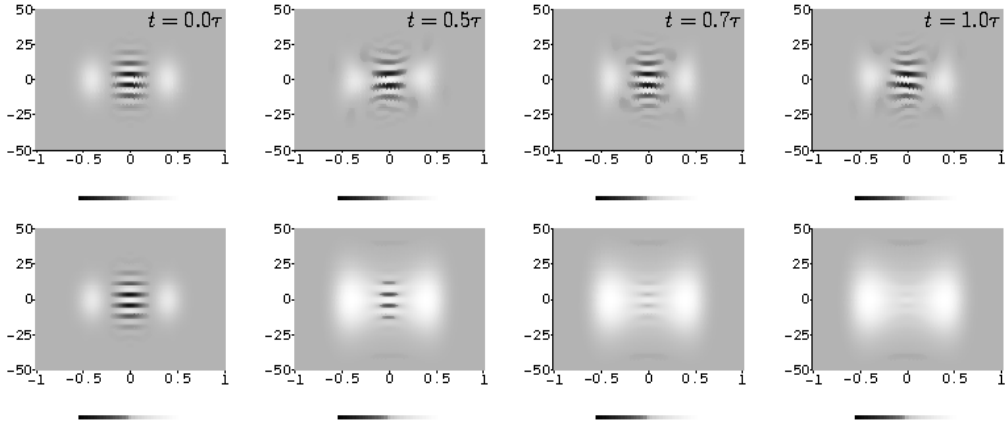,height=6cm,width=14cm,angle=0}}
\caption{Wigner functions for the isolated (row on top) and open (botton row)
systems, for the indicated times. The horizontal axis corresponds to $x$, vertical
axis to $p$. The medium grey shade on the background correspond to zero values for
the Wigner function, lighter and darker shades, respectively, to positive and
negative values of $W(x,p)$. The open system has totally decohered before the tunneling 
time-scale.}
\label{fig1}
\end{figure}

Fig.~\ref{fig1} shows the Wigner function of the system for the
indicated times, both for the isolated and open systems. The initial
Schr\"odinger cat state is depicted in the first picture of each row (left). Black 
fringes denote the 
quantum interference terms present in the initial state. Already at very early times,
the negative regions of $W(x,p)$ are considerably suppressed for the
open case when compared to the closed system, suggesting that
decoherence has a crucial role in the evolution. For $t \leq \tau$, 
$W(x,p)$ becomes positive definite and the open system displays
no tunneling interferences \cite{nuno}. Decoherence inhibits
tunneling. Once the interferences are destroyed, the spread of the Wigner function 
increases as a consequence of diffusion induced by the environment. As expected, 
decoherence has clearly been effective by the time 
the Wigner function is strictly positive every-where. As the tunneling time 
is reached, though the system is localized on the original wells, the Wigner 
function explores a large region of the phase space. On the contrary, for 
the closed system, negative values in the center of the phase space clearly indicate 
quantum behavior. The isolated system is tunneling from one to the 
other well.

\begin{figure}
\centerline{\psfig{figure=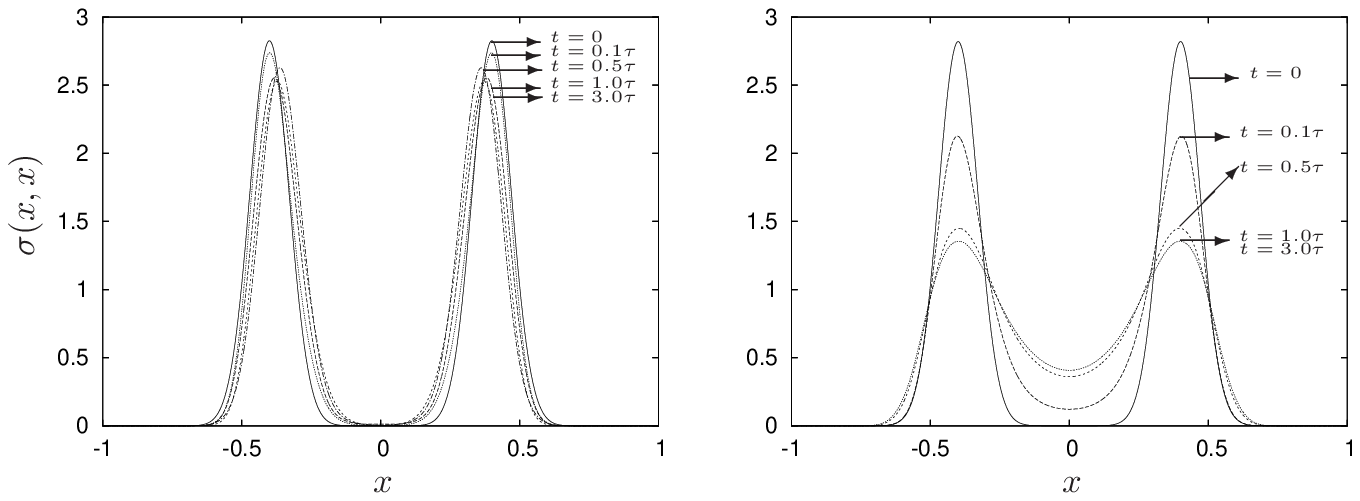,height=3.5cm,width=9cm,angle=0}}
\caption{Probability distribution for the isolated (left) and open
(right) systems for the indicated times.}
\label{fig2}
\end{figure}

In Fig.~\ref{fig2} we show the probability distribution $\sigma (x,x)$ for the isolated 
and open systems for the indicated times, measured in units of the estimated 
tunneling time $\tau$. It is worthy of note that for the 
closed case the state keeps its phase coherence and the tunneling 
effect is present. The state is localized on the original wells and the evolution 
is clearly unitary. In the open case, even though the system 
is still localized on the original wells for early times, the probability spreads due to the 
diffusion induced by the environment. Probability 
differs from the closed case owing to the noise-induced diffusion. $\sigma$ starts 
spreading for early times. By $t \sim 0.5\tau$ there is nonzero probability at the 
position of the barrier; clearly indicating that there is 
probability crossing the barrier but not by tunnel effect (as we have shown in the 
Wigner pictures, the system is already classical by the tunneling time). It is expected that 
the probability over the barrier grows in time, reaching an uniform value 
asymptotically. This fact  
suggests that there is a process of energy activation induced by the 
presence of the environment \cite{pre}.

When trying to interpret the post-decoherence behaviour of the open
system, several features of its dynamics should be kept in mind.
Firstly, one should emphasize that the initial condition is clearly
not the ground state of the composite system. As soon as the
interaction between the main system and the environment is turned
on, at $t=0$, the system will find itself in an excited state. In
relation to the new minima of the potential, the environment will
have a non-zero amount of energy. From a purely classical point of
view, this energy cannot be responsible for the excitation of the
particles over the potential barrier. In fact, the height of the
potential increases in relation to the new vacuum, in such a way 
that the {\it total} energy of the full system is still lower than
the barrier separating regions of positive and negative $x$. 
Note the fact that there are no fluctuations in the environment
 classically at T=0, plays a crucial role in this reasoning. Even
 for small but finite T, the energy of the environment would
 go as T. By choosing T small enough, this contribution could always
 be made smaller than the barrier height. As a consequence, and in contrast
with the high-T case, we will not
be able to describe the evolution of the quantum system after classicalization
by simply taking its classical exact equivalent. The
quantum fluctuations present in the initial state of the environment must
play a role in the ``activation''. One should note that these fluctuations are
not ``vacuum fluctuations" of the full system. The quantum nature of the environment,
which could be ignored at high-T, leads in this limit to important non-negligible
effects.
In terms of the master equation, the quantum fluctuations of the bath
oscillators generate non-zero $f(t)$ and $D(t)$ terms,
making diffusive phenomena possible. This is particularly true in the case of the anomalous
diffusion coefficient $f(t)$ that depends logarithmically on the cutoff
$\Lambda$ and thus can be considerably large \cite{pau}. Diffusion effects induced
by quantum fluctuations  are thus responsible for exciting the particles
over the potential barrier.
\begin{figure}
\centerline{\psfig{figure=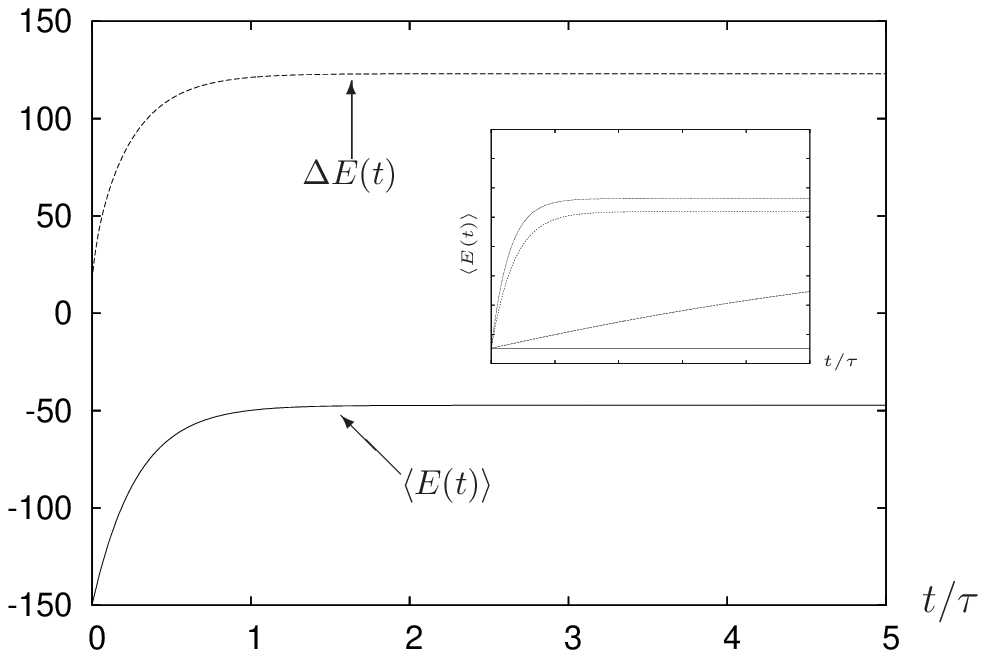,height=3.5cm,width=6cm,angle=0}}
\caption{Evolution in time of the mean energy of the system
and quantum dispersion of the energy for $\Lambda = 10 V_0$. In the inset, we 
show the dependence of 
the energy with the cutoff $\Lambda$.
On botton, the smallest value of $\Lambda$; straight dashed line is $\Lambda = V_0/10$.
Dotted lines on top of the inset, are larger values of the cutoff: $\Lambda = 10 V_0$,
and $\Lambda = V_0$ just below.}
\label{fig3}
\end{figure}
The interplay of decoherence and excitation processes in
the $T=0$ case deserves a deeper insigh. For both quantities, the value of the environment
frequency cutoff $\Lambda$, seems to play an important role,
affecting both the decoherence time, and the excitation process in
the same direction. Since we do not have explicit estimates of the
``activation'' time in terms of $\Lambda$, it is hard to predict
whether there is a regime for which decoherence happens fast enough
and excitation is considerably delayed. Numerical results presented
in Fig. \ref{fig3} suggests that this is not possible. The figure
show the mean energy and its dispersion as a function of time. Also
in the inset, we show the energy of the main
system for several choices of the cutoff. $\Lambda$ varies from the
smallest frequency present in the system; i.e. the difference
between the first exited and the ground state energy levels, $E_0 - E_E$; and
 $\Lambda = 10 V_0$. Also shown are two intermediate cutoff values, $\Lambda = V_0/10$ and
 $\Lambda = V_0$. By lowering $\Lambda$,
the ``activation'' time is indeed
postponed, but so is decoherence \cite{pau}. In this situation the particles are simply
able to tunnel back and forth the two minima for a longer period. Higher values
of the cutoff, on the other hand lead to both fast decoherence and fast
``activation''. As a result we were never able to localize the particle
on one of the wells, with tunneling and ``activation'' being simultaneously
suppressed. The dispersion in the energy distribution $\Delta E(t)$ shows 
that fluctuations in the environment at $T = 0$ are non-neglegible for large 
values of $\Lambda$. The latter have a faster growth in time and reach a bigger final 
value than the mean energy of the system. Noise-induced energy activation is a 
consequence of this fact.

We would like to thank N. Antunes for his collaboration
on this project. We also thanks B.L. Hu for usefull comments,
and H. Thomas Elze for the organization of DICE'06. This work is supported by
UBA, Conicet, and ANPCyT Argentina.


\section*{References}
\begin{thebibliography}{[99]}

\bibitem{Shin04} Shin Y, Saba M, Pasquini T A, Ketterle W, Pritchard D E, and Leanhardt 
A E 2004, Phys. Rev. Lett. {\bf 92}, 050405

\bibitem{Anderson02} Anderson E, Calarco T, Folman R, Anderson M, Hessmo B, and 
Schmiedmayer 2002, Phys. Rev. Lett. {\bf 88}, 100401

\bibitem{Shin05} Shin Y, Jo  G B, Saba M, Pasquini T A, Ketterle W, and Pritchard D E 
2005 Phys. Rev. Lett. {\bf 95}, 170402

\bibitem{Menotti01} Menotti C, Anglin J R, Cirac J I, and Zoller P  2001 Phys. Rev. {\bf A63}, 
023601

\bibitem{caleg} Caldeira A O and Leggett A J 1981 Phys. Rev. Lett.
{\bf 46}, 211; Leggett A J, Chakravarty S, Dorsey A T, Fisher M P A, 
Garg A, and Zwerger W 1987 Rev. Mod. Phys. {\bf 59}, 1


\bibitem{takagi} Takagi S, {\it Macroscopic Quantum Tunneling}, Cambridge University Press,
Cambridge, UK (2002).

\bibitem{hanggi} H\"anggi P, Talkner P, and Borkovek M 1990 Rev. Mod. Phys. {\bf 62},
251

\bibitem{Coleman} Coleman S, {\it Aspects of Symmetry}; Cambridge University Press,
NY (1985).

\bibitem{pau} Lombardo F C and Villar P I 2005 Phys. Lett. {\bf B336}, 16 

\bibitem{calzetta1} Calzetta E and Verdaguer E 2006 J.Phys.{\bf A}39, 9503 

\bibitem{pre} Antunes N, Lombardo F C, Monteoliva D and Villar P I 2006 Phys. Rev. {\bf E73},
066105 

\bibitem{frazer} Dounas-Frazer D R, Hermundstad A M and Carr L D quant-ph/0609119; 
Dounas-Frazer D R and Carr L D quant-phys/0610166

\bibitem{pitaevskii} Pitaevskii L and Stringari S 2001, Phys. Rev. Lett. {\bf 87}, 
180402


\bibitem{diana} Lombardo F C, Mazzitelli F D, and
Monteoliva D 2000 Phys. Rev. D {\bf  62}, 045016 

\bibitem{nuno} Antunes N, Lombardo F C and Monteoliva D 2001 Phys. Rev. {\bf E64},
066118




\end{thebibliography}
\end{document}